\documentclass[prd,twocolumn,a4paper]{revtex4}

\usepackage{amsmath}
\usepackage{longtable}

\providecommand{\be}{\begin{equation}}
\providecommand{\ee}{\end{equation}}
\providecommand{\bea}{\begin{eqnarray}}
\providecommand{\eea}{\end{eqnarray}}

\begin{document}

\title{Does a relativistic metric generalization of Newtonian
gravity exist in 2+1 dimensions?}
\date{October 23, 2002}
\author{J.L. Alonso, J.L. Cort\'es, and V. Laliena}
\email{buj@gteorico.unizar.es,cortes@leo.unizar.es,laliena@posta.unizar.es}
\affiliation{Departamento de F\'{\i}sica Te\'orica, Universidad de
Zaragoza, C. Pedro Cerbuna 12, E-50009 Zaragoza (Spain)}

\begin{abstract}
It is shown that, contrary to previous claims, 
a scalar tensor theory of Brans-Dicke type
provides a relativistic generalization of Newtonian gravity 
in 2+1 dimensions. The theory is metric and 
test particles follow the space-time geodesics. 
The static isotropic solution is studied in vacuum and in regions
filled with an incompressible perfect fluid. It is shown that the
solutions can be consistently matched at the interface matter-vacuum,
and that the Newtonian behavior is recovered in the weak field regime. 
\end{abstract}

\maketitle

\section{\label{sec:intro} Introduction}

Newtonian gravity, a theory of gravitational phenomena which is
invariant under Galilean transformations and, therefore, valid only in
the low energy (weak fields and slow motions) regime, must be
generalized to a Relativistic Theory of Gravitation (RTG). 
Einstein General Relativity (EGR) is a good candidate for RTG in 3+1
dimensions, but other possibilities, as Brans-Dicke theory (BDT)
\cite{BD}, have been 
proposed. Indeed, it is believed by many that a quantum theory of
gravitation, which seems unavoidable if we want to deal with
gravitational phenomena at the Planck scale, must contain something
more than ERG \cite{GSW}.

In 2+1 dimensions, EGR is not a RTG. The
Riemann-Christoffel tensor is uniquely determined by the Ricci tensor,
which vanishes outside the sources. Hence, space-time is flat in
regions devoid of matter, the geodesics are straight lines and 
test particles do not feel any gravitational field \cite{deser}. 
A proper RTG in 2+1 dimensions
needs some additional ingredient besides the metric tensor of EGR
\cite{teleparallel}. 
A minimal candidate for RTG is a scalar tensor theory of Brans-Dicke
type. It has been claimed that even theories of this sort, which are
much richer than EGR, do not describe Newtonian gravity in the low
energy limit \cite{barrow}. 
We will show in this paper that BDT in 2+1 dimensions
reproduces Newtonian gravity when the low energy regimen is
consistently analysed. The additional ingredient that seems necessary
to construct a quantum theory of gravitation in 3+1 dimensions appears
at an earlier stage in lower dimensions.

The construction of a RTG in dimensions lower than 3+1 is interesting
because it may allow to study phenomena characteristic of gravity,
which have 3+1 dimensional analogues,
such as gravitational instabilities and black holes, in a
simplified context \cite{lowgrav,BD2D}. 
 
The paper is organized as follow. In section II the 
equations of scalar-tensor theories in $D$-dimensional space-time are derived,
the particular case of BDT is identified, and the peculiarities of
$D=3$ and $D=2$ are discussed. Section III is devoted to the study of
the weak field limit and in section IV the fields corresponding to
static isotropic extended sources in $D=3$ are analysed. The solutions inside
and outside the source can be consistently matched and 
reproduce the Newton gravitational field in the weak field limit.
Finally, section V contains some concluding remarks.

\section{\label{sec:equations} Scalar-tensor theories in
D--dimensions}

The minimal modification of EGR introduces, besides the
metric tensor, a scalar field associated to the gravitational
interaction. The most natural way to interpret it is noting that it sets
the strength of the gravitational coupling at each space-time point,
which is thus determined by the mass distribution of the universe. It has
the advantage that it incorporates Mach principle better than EGR.
The fact that the scalar field is related to the gravitational
interaction strength implies that its coupling to the metric tensor
cannot be minimal: it must be of the type scalar--tensor
interaction. The action for the gravitational sector in
such kind of theories in $D$-dimensional space-time is  

\begin{equation}
{\cal S}_g \;=\; - \int d^D x \sqrt{g} \left[\phi R + W(\phi) 
 g^{\mu\nu} \phi_{;\mu} \phi_{;\nu}\right] \, ,
\label{Sg}
\end{equation} 

\noindent
where $g_{\mu\nu}$ is the metric tensor, whose signature is taken
$(-,+,\ldots,+)$, $g=-\det g_{\mu\nu}$, $R$ is the scalar
of curvature, and $\phi$ is the scalar field that acts as the
inverse gravitational coupling at each space-time point.
The action contains an unspecified function of $\phi$, $W$.
Different choices of such function give different scalar tensor
theories.  

The coupling of the matter to gravity depends only on the metric
tensor in a covariant way, and does not depend of the scalar field.
Hence, the variation of the matter action, ${\cal S}_m$, under
variations of the metric, $\delta g_{\mu\nu}$, can be written as
$\delta {\cal S}_m=1/2 \int d^Dx\sqrt{g}T^{\mu\nu}\delta g_{\mu\nu}$,
and general covariance implies a continuity equation for $T^{\mu\nu}$:

\begin{equation}
T^{\mu\nu}_{;\nu}\;=\;0 \, , \label{cont}
\end{equation}

\noindent
that describes the exchange of energy between matter and
gravity. Furthermore, this equation ensures that test particles move
along the geodesics of space-time associated to the metric
$g_{\mu\nu}$ and thus the equivalence principle is preserved. In this
sense, scalar-tensor theories are metric theories of gravitation.

The equations for the metric tensor and the scalar field that follow
from the action 
${\cal S}={\cal S}_g+{\cal S}_m$  are

\begin{widetext}

\begin{subequations}
\label{eq1}
\begin{eqnarray}
\phi \left[R_{\mu\nu} - \frac{1}{2} g_{\mu\nu} R\right] 
+
(g_\mu^\rho g_\nu^\sigma-g_{\mu\nu}g^{\rho\sigma})\phi_{;\rho;\sigma}
+
W(\phi) \left[g_\mu^\rho g_\nu^\sigma
-\frac{1}{2}g_{\mu\nu}g^{\rho\sigma}\right] \phi_{;\rho} \phi_{;\sigma}
&=& -\frac{1}{2} T_{\mu\nu}  \label{eqg} \\
R \:-\: g^{\rho\sigma} \left[2 W(\phi) \phi_{;\rho;\sigma} +  
W'(\phi) \phi_{;\rho} \phi_{;\sigma}\right] &=& 0 \, ,
\label{eqphi}
\end{eqnarray}
\end{subequations}

\noindent
where the prime means derivation with respect to $\phi$.

For $D\neq 2$, Eq.~(\ref{eqg}) allows to express $R$ as a function of
$\phi$ and its covariant derivatives and the trace of the
energy-momentum tensor, $T=g_{\mu\nu}T^{\mu\nu}$. Thus,
Eqs.~(\ref{eq1}) can be written in the following
form:

\begin{subequations}
\label{eq2}
\begin{eqnarray}
&& \phi R_{\mu\nu} \;=\; 
- \left[\frac{1}{D-2} g_{\mu\nu} 
g^{\rho\sigma} +  g_\mu^\rho g_\nu^\sigma\right] \phi_{;\rho;\sigma} -
W(\phi) \phi_{;\mu} \phi_{;\nu} -
\frac{1}{2} \left[T_{\mu\nu} - \frac{1}{D-2} g_{\mu\nu} T\right] \, ,
\label{eqg2}  \\
&& 2\left[\frac{D-1}{(D-2)\phi} + W(\phi)\right] g^{\rho\sigma}
\phi_{;\rho;\sigma} + \left[\frac{W(\phi)}{\phi} + W'(\phi)
\right] g^{\rho\sigma} \phi_{;\rho} \phi_{;\sigma} 
- \frac{1}{(D-2)\phi}\,T \;=\; 0 \, .
\label{eqphi2}
\end{eqnarray}
\end{subequations}

\end{widetext}

The above equations are particularly simple with the choice
$W(\phi)=\omega/\phi$, where $\omega$ is a constant. In this case
we get the field equations of Brans-Dicke theory \cite{BD}:

\begin{subequations}
\label{eq3}
\begin{eqnarray}
&& g^{\rho\sigma} \phi_{;\rho;\sigma} 
= \frac{\lambda}{2 (1+\omega)}\,T\, ,
\label{eqphi3} \\
&& \phi R_{\mu\nu} = 
- \left[
\phi_{;\mu;\nu} + \frac{\omega}{\phi} \phi_{;\mu} \phi_{;\nu}
\right] - \frac{1}{2} \left[T_{\mu\nu} - 
\lambda g_{\mu\nu} T\right] \, . \hspace*{0.75truecm}
\label{eqg3}
\end{eqnarray}
\end{subequations}

\noindent
where $\lambda$ is a function of $\omega$ and $D$:

\begin{equation}
\lambda\;=\;\frac{1\,+\,\omega}{(D-1)\:+\:\omega(D-2)} \, .
\end{equation}

Introducing the parametrisation $\phi=\frac{1}{\cal G}\mathrm e^\xi$
(then $\phi_{;\mu}=\phi\xi_{;\mu}$ and 
$\phi_{;\mu;\nu}=\phi(\xi_{;\mu;\nu}+\xi_{;\mu}\xi_{;\nu})$), where
$\xi$ is dimensionless and the constant ${\cal G}$ appears for
dimensional reasons, the
equations take the form

\begin{subequations}
\label{eqRxi}
\begin{eqnarray}
&&\hspace*{-1truecm} 
g^{\rho\sigma} \left(\xi_{;\rho;\sigma} + \xi_{;\rho} 
\xi_{;\sigma}\right) = \frac{{\cal G}}{2}  \mathrm e^{-\xi}  
\frac{\lambda}{1+\omega} T \, , \label{eqxi} \\
&& \hspace*{-1truecm} R_{\mu\nu}=-\xi_{;\mu;\nu} - 
(1+\omega) \xi_{;\mu} \xi_{;\nu} - 
\frac{{\cal G}}{2} \mathrm e^{-\xi} \left[T_{\mu\nu}- 
\lambda g_{\mu\nu} T\right] . \hspace{0.25truecm}
\label{eqR}
\end{eqnarray}
\end{subequations}

\noindent
Such parametrisation assumes only that the scalar field
has a definite sign and, as we shall see, is appropiate to
analyse the existence of the Newtonian limit.

The case $D=2$ is especial, since 
$R_{\mu\nu}=\frac{1}{2}g_{\mu\nu} R$, and the equations 
read

\begin{subequations}
\begin{eqnarray}
&&g^{\rho\sigma} \left(\xi_{;\rho;\sigma} + \xi_{;\rho} 
\xi_{;\sigma}\right) \;=\; \frac{{\cal G}}{2} \, e^{-\xi} \, T \, ,  \\
&&R \;=\; \omega \, \left( {\cal G} e^{-\xi} T \:-\: g^{\rho\sigma}
\xi_{;\rho} \xi_{;\sigma}\right) \, .
\end{eqnarray}
\end{subequations}

\section{\label{sec:wfl} Perturbative solution in D-dimensions}

If the energy-momentum tensor of matter, $T^{\mu\nu}$, vanishes, 
the system of equations~(\ref{eqRxi}) have the
vacuum solution $g_{\mu\nu}=\eta_{\mu\nu}$ and $\xi=0$, where
$\eta_{\mu\nu}$ is the Minkowski metric tensor. This solution fixes
the arbitrariness in the coordinate system and definition of ${\cal G}$.
For a weak $T_{\mu\nu}$ the solution of 
Eqs.~(\ref{eqRxi}) can be obtained perturbatively
as a series in powers of ${\cal G}$:

\begin{subequations}
\begin{eqnarray}
\xi(x) &=& \sum_{n=1}^{\infty} {\cal G}^n \xi^{(n)}(x)\, ,
\label{xin} \\
g_{\mu\nu}(x) &=& \eta_{\mu\nu} \:+\: \sum_{n=1}^{\infty} {\cal G}^n 
g_{\mu\nu}^{(n)}(x) \, , 
\label{gn} \\
T_{\mu\nu}(x) &=& T_{\mu\nu}^{(0)}(x)\:+\:\sum_{n=1}^\infty {\cal G}^n
T_{\mu\nu}^{(n)}(x) \, ,
\label{Tn}
\end{eqnarray}
\end{subequations}

\noindent
where $T_{\mu\nu}^{(0)}$ is the energy-momentum tensor of matter in
the absence of gravitation. Imposing Eqs.~(\ref{eqRxi}) and the
continuity equation~(\ref{cont}) order by order
in ${\cal G}$ leads to a set of linear
equations for $\xi^{(n)}$, $g_{\mu\nu}^{(n)}$, and
$T_{\mu\nu}^{(n)}$. 
To first order in ${\cal G}$, Eqs.~(\ref{eqRxi}) yield

\begin{subequations}
\begin{eqnarray}
&& \eta^{\rho\sigma} \frac{\partial^2 \xi^{(1)}}{\partial x^\rho
\partial x^\sigma} \;=\;  
\frac{\lambda}{2\,(1+\omega)} \, T^{(0)} \, ,
\label{xi1} \\
&& R_{\mu\nu}^{(1)} \;=\; 
- \frac{\partial^2 \xi^{(1)}}{\partial x^\mu
\partial x^\nu} \:-\: \frac{1}{2} \left[T_{\mu\nu}^{(0)} - 
\lambda \eta_{\mu\nu} T^{(0)}\right] \, .\hskip 0,3cm 
\label{R1}
\end{eqnarray}
\end{subequations}

For a static field produced by non-relativistic matter 
($T_{\mu\nu}^{(0)}=0$ for $(\mu,\nu)\neq(0,0)$ and
$T_{00}^{(0)}=\rho$, where $\rho$ is the density of matter),
$R_{00}^{(1)}=\frac{1}{2} \nabla^2 g_{00}^{(1)}$. 
In this case, for $D>2$ we have:

\begin{equation}
\nabla^2 g_{00}^{(1)}\;=\;-(1\,-\,\lambda)\,\rho \, , 
\label{swf}
\end{equation}

\noindent
and for $D=2$:

\begin{equation}
\nabla^2 g_{00}^{(1)}\;=\;-\frac{\omega}{4}\,\rho \, .
\label{swf2d} 
\end{equation}

The Newtonian potential is identified from the geodesic
equation as $V_\mathrm{N}=-\frac{\cal G}{2}g_{00}^{(1)}$. Remembering that the
Newtonian potential in $D-1$ spatial dimensions verifies the 
Poisson equation 

\begin{equation}
\nabla^2 V_\mathrm{N}\;=\;S_{D-2}\,G_\mathrm{N}\,\rho \, ,
\end{equation}

\noindent
where $S_n$ is the area of the n-dimensional unit sphere \cite{foot} 
and $G_\mathrm{N}$ the Newton constant, we have for $D>2$

\begin{equation}
G_\mathrm{N}\;=\;\frac{\cal G}{2S_{D-2}}\,(1\,-\,\lambda)
\, , \label{ncd}
\end{equation}

\noindent
and for $D=2$

\begin{equation}
G_\mathrm{N}\;=\;\frac{{\cal G}\omega}{4} \, . \label{nc2d}
\end{equation}

\noindent
As expected, we see that $G_\mathrm{N}$ vanishes as $\omega\rightarrow\infty$
in $D=3$, and it tends to ${\cal G}/16\pi$ in $D=4$.

Note that the Newtonian limit of the scalar-tensor with a
coupling as given in Eqs.~(\ref{ncd})~and~(\ref{nc2d}) is 
compatible with 
a variable effective gravitational coupling proportional to the
reciprocal of the scalar field, 
$G_\mathrm{eff}=\frac{1-\lambda}{2S_{D-2}}\frac{1}{\phi}$.
In fact, $G_\mathrm{N}$ is just the first order term in an expansion
of $G_\mathrm{eff}$ in powers of ${\cal G}$.

\section{\label{sec:sf} Field of static isotropic sources in D=3}

The perturbative expansion is rather formal and it is interesting to see
that it is consistent in the most important case of static
and isotropic sources. To this end,
let us compute the field produced by a static and isotropic source in
regions devoid of sources in $D=3$. 
In a suitable reference system, the metric
can be written in the standard form:

\begin{equation}
d\tau^2\;=\;\mathrm e^{2\beta}\,dt^2\:-\:\mathrm e^{2\alpha}\,dr^2\:
-\:r^2\,d\varphi^2 \, ,
\end{equation}

\noindent
where $\alpha$ and $\beta$ are functions of $r$ alone.
In vacuum ($T_{\mu\nu}=0$) Eqs.~(\ref{eqRxi}) yield

\begin{subequations}
\label{solvac}
\begin{eqnarray}
\xi &=& \xi_0\:+\:C_\xi\,\ln (r/r_0) \, ,  \\
\alpha &=& \alpha_0\:+\:C_\alpha\,\ln(r/r_0) \, , \\
\beta &=& \beta_0\:+\:C_\beta\,\ln(r/r_0)\, ,
\end{eqnarray}
\end{subequations}

\noindent
where $r_0$ is some arbitrary length scale, $\xi_0$, $\alpha_0$, and 
$\beta_0$ and $C_\xi$ are constants, and

\begin{subequations}
\begin{eqnarray}
C_\alpha &=& \frac{1}{2}\frac{(\omega+2) C_\xi^2}{1+C_\xi} \, , \\
C_\beta &=& \frac{1}{2}\frac{\omega C_\xi^2-2 C_\xi}{1+C_\xi} \, . 
\label{eqcb} 
\end{eqnarray}
\end{subequations}

The fields are singular at the origin and must be produced by some
source extended about (or concentrated at) the origin.
The constant $C_\xi$ is determined by the source as follows:
multiplying Eq.~(\ref{eqphi3}) by $\sqrt{g}$ and integrating in
$drd\varphi$, and assuming that a static and isotropic source is confined
within a finite spatial region, we get

\begin{equation}
C_\xi\;=\;-\exp(\alpha_0-\beta_0-\xi_0)\,
\frac{\cal G}{4\pi(\omega+2)}\,M\, ,
\label{cxi}
\end{equation} 

\noindent
where the mass, $M$, is defined as $M=\int drd\varphi\sqrt{g} T$.

In the weak field regime ${\cal G}M$ is small. 
Eqs. (\ref{xin})-(\ref{gn}) imply that $\xi_0$, $\alpha_0$, and 
$\beta_0$ are proportional to ${\cal G}$. 
Then to first order in ${\cal G}$ we have

\begin{equation}
\mathrm e^{2\beta}\;\approx\;1+2\frac{{\cal G}}{4\pi(2+\omega)}M\,
\ln\frac{r}{r_0} \, .
\end{equation}

\noindent
A comparison with the Newtonian solution, 
$V_\mathrm{N}=G_\mathrm{N} M\ln(r/r_0)$,
gives the Newton constant $G_\mathrm{N}=\mathcal{G}/[4\pi(2+\omega)]$, 
the same
obtained from field equations in presence of matter in the weak field 
regime.

Eqs. (\ref{eqRxi}) do not admit solutions with
a static point mass as source \cite{barrow}. Indeed, they do not admit
static and isotropic solutions with concentrated sources 
(energy-momentum tensor that contains Dirac deltas). The reason is
that it is not possible to find solutions $\alpha$, $\beta$, and $\xi$
of the sourceless equations with
singularities cancelling the source singularities in the field equations.
We are thus lead to consider extended sources. This is interesting
because doubts have
been cast on the existence of static and isotropic solutions of
Eqs.~(\ref{eqRxi}) even with extended sources \cite{barrow}. 
To investigate this in depth,
let us consider as source an incompressible fluid of density
$\rho$ confined on a disk of radius $r_\mathrm{M}$. The corresponding
energy-momentum tensor is

\begin{equation}
T_{\mu\nu}\;=\;p\,g_{\mu\nu}\:+\:(\rho+p) U_\mu U_\nu\, ,
\end{equation}

\noindent
where $p$ is the pressure and $U_\mu$ the four velocity, which
verifies $g^{\mu\nu}U_\mu U_\nu=-1$. For a static fluid we have

\begin{equation}
T_{tt}\;=\;\mathrm e^{2\beta}\,\rho\: ,
\hspace{0.5cm} T_{rr}\;=\;\mathrm e^{2\alpha}\,p\: , \hspace{0.5cm}
T_{\varphi\varphi}\;=\;r^2\,p\: ,
\end{equation}

\noindent
and the remaining components vanish. Covariant conservation of 
$T_{\mu\nu}$ implies the equation of hydrostatic equilibrium 
\cite{weinberg}:

\begin{equation}
\beta'\;=\;-\frac{p'}{p+\rho}\, ,
\label{hydros}
\end{equation}

\noindent
where the prime indicates derivation respect to $r$. Using the
dimensionless variable $\hat{r}=r/r_\mathrm{M}$ and $\bar{p}=p/\rho$, 
and the dimensionless parameter 
$\kappa={\cal G}r_\mathrm{M}^2\rho/[2 (2+\omega)]$, 
Eqs.~(\ref{eqRxi}) read:

\begin{widetext}
\begin{subequations}
\label{sieq} 
\begin{eqnarray}
\xi^{\prime\prime}\:+\:\frac{1}{\hat{r}}\xi'\:+\:\xi'\,(\beta'+\xi'-\alpha') 
&=& -\kappa\,(1-2\bar{p})\,\mathrm e^{2\alpha-\xi} \, , \\
\beta^{\prime\prime}\:+\:\frac{1}{\hat{r}}\beta'\:+\:
\beta'\,(\beta'+\xi'-\alpha') &=& 
\kappa\,[\,1+2(1+\omega)\bar{p}\,]\,\mathrm e^{2\alpha-\xi} \, , \\
\beta^{\prime\prime}\:-\:\frac{1}{\hat{r}}\alpha'\:+\:\xi^{\prime\prime}\:+\:
\beta^{\prime 2}\:+\:(1+\omega)\xi^{\prime 2}
-\alpha'(\beta'+\xi') &=& 
-\kappa\,[\,1+\omega-\omega\bar{p}\,]\,\mathrm e^{2\alpha-\xi} \, ,\\
\frac{1}{\hat{r}}\,(\beta'+\xi'-\alpha') &=&
-\kappa\,[\,1+\omega-\omega\bar{p}\,]\,\mathrm e^{2\alpha-\xi} \, , 
\end{eqnarray}
\end{subequations}
\end{widetext}

\noindent
where now the primes stand for derivation respect to $\hat{r}$.

Notice that there is no static and isotropic solution for
dust ($p=0$), since then $\beta'=0$ and this forces $\kappa=0$, i.e.,
$\rho=0$. 

With nonzero $\kappa$, the system of Eqs.~(\ref{sieq}), supplemented
with~(\ref{hydros}), is very complicated. For
$\kappa=0$ there is a trivial solution : $\xi=\alpha=\beta=\bar{p}=0$.
For $\kappa\neq 0$ the solution can be found iteratively, 
by an expansion in powers of $\kappa$:


\begin{equation}
\begin{array}{ll}
\xi \;=\; \sum_{i=1}^\infty\,\kappa^i\,\xi_i \, , \hspace{0.5truecm} &
\alpha \;=\; \sum_{i=1}^\infty\,\kappa^i\,\alpha_i \, ,\vspace{0.2truecm} \\
\beta \;=\; \sum_{i=1}^\infty\,\kappa^i\,\beta_i \, ,
\hspace{0.5truecm} & 
\bar{p} \;=\; \sum_{i=1}^\infty\,\kappa^i\,\bar{p}_i \, .
\end{array}
\end{equation}

At order $\kappa^i$ the equations for $\xi_i$, $\alpha_i$, and
$\beta_i$ are linear and very easy to solve. The solutions, however,
become more and more cumbersome as the order
increases. Table~\ref{tab:sol} displays the solution up to order
$\kappa^4$, with the boundary conditions discussed below.
With the aid of programs of symbolic calculus one can easily find
the solution up to very large orders.

Boundary conditions are fixed by requiring that $\alpha$,
$\beta$, and $\xi$ be regular at the origin. This eliminates the
singular solutions, which contain $\ln\hat{r}$, and fixes two
integration constants. The ambiguity in the choice of a reference system
and a scale for the gravitational interaction (${\cal G}$ has been 
arbitrarily chosen) is reflected in the presence of three more
integration constants. They can be fixed, for instance, by requiring
that at the origin the coordinate system be locally inertial and 
$\xi$ vanish.
In addition, we require that the pressure vanishes at
$\hat{r}=1$. In this way, the solution is completely determined.

Let us analyze the matching with the vacuum solution (\ref{solvac})
for $\hat{r}>1$. It is convenient to choose as length scale 
$r_0=r_\mathrm{M}$.
The boundary conditions at $\hat{r}=1$ are the continuity of $\alpha$, $\beta$,
$\xi$, $\beta'$, and $\xi'$. Since $T_{\mu\nu}$ has a jump at
$\hat{r}=1$, it follows that $\beta^{\prime\prime}$, $\xi^{\prime\prime}$, 
and $\alpha'$ must
also have a jump. The constants $\alpha_0$, $\beta_0$, and $\xi_0$
must be chosen to guarantee the continuity of $\alpha$, $\beta$, and
$\xi$. For the derivatives we have:

\begin{subequations}
\begin{eqnarray}
\sum_{i=1}^\infty\,\kappa^i\,\xi'_i(\hat{r}=1) &=& C_\xi \, , \\
\sum_{i=1}^\infty\,\kappa^i\,\beta'_i(\hat{r}=1) &=& C_\beta \, .
\end{eqnarray}
\end{subequations}

\noindent
Although $C_\xi$ and $C_\beta$ are linked by Eq.~(\ref{eqcb}), the above
two equations are simultaneously satisfied since Eqs.~(\ref{sieq}) 
guarantee that the following relation holds at the points where the 
pressure vanishes:

\begin{equation}
2 \beta'\,\left(\frac{1}{\hat{r}}\,+\,\xi'\right)
\;=\;\omega\xi^{\prime 2}\:-\:
\frac{2}{\hat{r}}\xi' \, .
\end{equation}

\noindent
Thus, the condition that pressure vanishes at $\hat{r}=1$ is not only
physically sound, but necessary to match with the vacuum solution.

The matching can be made order by order in $\kappa$. In such case we
have 

\begin{equation}
C_\xi\;=\;\sum_{i=1}^\infty\,\kappa^i\,C_\xi^{(i)} \, .
\end{equation}

\noindent
As a check, it can be seen that Eq.~(\ref{cxi}) is verified. 

The solution up to order $\kappa^4$ is displayed in 
Table~\ref{tab:sol}.
The fields $\beta$ and $\alpha$ are positive and monotonically
increasing order by order, and 
the preasure is positive and monotonically decreasing order by order, 
vanishing at the border of the mass distribution ($\hat{r}=1$).
This is in agreement with intuition, since a positive preasure is
required to balance the gravitational attraction. Gravitational
collapse is impossible since the fluid is incompressible.

Up to order $\kappa^4$, the limit $\omega\rightarrow\infty$ 
gives (recall that $\kappa\sim 1/\omega$) 
$\xi=\beta=\bar{p}=C_\xi=0$ and

\begin{equation}
\alpha=\frac{1}{2}\bar{\kappa}\hat{r}^2
+\frac{1}{4}\bar{\kappa}^2\hat{r}^4+
\frac{1}{6}\bar{\kappa}^3\hat{r}^6
+\frac{1}{8}\bar{\kappa}^4\hat{r}^8\, ,
\label{appwinf}
\end{equation}
 
\noindent
where $\bar{\kappa}={\cal G} r_\mathrm{M}^2\rho/2$.
On the other hand, we can take the limit $\omega\rightarrow\infty$
of Eqs.~(\ref{sieq}). This forces $\xi'=0$ and then the solution
with the boundary conditions previously described 
is $\xi=\beta=\bar{p}=0$ and 

\begin{equation}
\alpha\;=\;-\frac{1}{2}\,\ln(1\,-\,\bar{\kappa}\hat{r}^2) \, .
\label{exwinf} 
\end{equation}

\noindent
We see that Eq.~(\ref{appwinf}) is the Taylor expansion of the exact
solution~(\ref{exwinf}) around $\bar{\kappa}\hat{r}^2=0$.
Since $\hat{r}\leq 1$, the expansion will converge 
for $\bar{\kappa}<1$. For finite $\omega$, the convergence radius
of the expansion in powers of $\kappa$ is an open question.

We can understand that the preasure is zero in the
$\omega\rightarrow\infty$ limit by
recalling that EGR in 2+1 dimensions produces no force between
separated sources. If we represent a continuous mass distribution
as a continuum limit of localized sources, we understand that the 
preasure necessary to balance gravity is zero.

The field of a static isotropic source in Brans-Dicke theory in $D=2$
has been obtained in \cite{BD2D}. It can be seen that the
Newton constant obtained from it agrees with Eq.~(\ref{nc2d}).

\begin{table*}[t!]
\begin{tabular}{ll}
\hline\hline
$\xi_1$ \ \ & $-\frac{1}{4}\,\hat{r}^2$ \\
\hline
$\xi_2$ \ \ \ \ & $\frac{1}{8}\,\hat{r}^2\:-\:
\frac{1}{64}\,(9+6\omega) \hat{r}^4$ \\
\hline
$\xi_3$ & $\frac{1}{128}\,\left[(15+12\omega)\hat{r}^2+(1+5\omega)\hat{r}^4
-\frac{1}{6}(40 \omega^2+114 \omega + 71)\hat{r}^6\right]$ \\
\hline
$\xi_4$ & $\frac{1}{192}(26+38\omega+15\omega^2)\hat{r}^2\:+\:
\frac{1}{2048}(31+71\omega+60\omega^2)\hat{r}^4
\:+\:\frac{1}{1536}(32\omega^2+29\omega-27)\hat{r}^6$ \\
 & $\:-\: 
\frac{1}{24576}(1710+4457\omega+3508\omega^2+840\omega^3)\hat{r}^8$ \\
\hline\hline
$\beta_1$ & $\frac{1}{4}\,\hat{r}^2$ \\
\hline
$\beta_2$ & $\frac{1}{8}\,(1+\omega)\,\hat{r}^2\:+\:
\frac{1}{64}\,(5+4\omega)\,\hat{r}^4$ \\
\hline
$\beta_3$ & $\frac{1}{128}\,\left[(12 \omega^2+27 \omega+15) \hat{r}^2\:+\:
(4 \omega^2+7 \omega + 7) \hat{r}^4
\:+\:\frac{1}{6}(24 \omega^2+64 \omega+35) \hat{r}^6\right]$ \\
\hline
$\beta_4$ & $\frac{1}{192}(26+64\omega+53\omega^2+15\omega^3)\hat{r}^2\:+\:
\frac{1}{2048}(121+189\omega+128\omega^2+48\omega^3)\hat{r}^4$ \\
 & $+\:\frac{1}{1536}(63+109\omega+76\omega^2+24\omega^3)\hat{r}^6 \:+\:
  \frac{1}{24576}(802+2291\omega+1932\omega^2+480\omega^3)\hat{r}^8$ \\
\hline\hline
$\alpha_1$ & $\frac{1}{2}(1+\omega)\,\hat{r}^2$ \\
\hline
$\alpha_2$ & $\frac{1}{4}\,\hat{r}^2\:+\:
\frac{1}{32}(8 \omega^2+19 \omega+8) \hat{r}^4$ \\
\hline
$\alpha_3$ & $\frac{1}{64} (12 \omega+15) \hat{r}^2\:+\:
\frac{5}{32}(\omega+1) \hat{r}^4\:+\:
\frac{1}{384}(64 \omega^3+238 \omega^2+251 \omega+73) \hat{r}^6$ \\
\hline
$\alpha_4$ & $\frac{1}{96} (15 \omega^2+38 \omega +26) \hat{r}^2\:+\:
\frac{1}{512}(60 \omega^2+123 \omega+83) \hat{r}^4\:+\:
\frac{1}{256} (35 \omega^2+84 \omega+39) \hat{r}^6$ \\
 & $+\:\frac{1}{12288}
(1536 \omega^4+7756 \omega^3+13232 \omega^2 + 8933 \omega +2002) \hat{r}^8$ \\
\hline\hline
$\bar{p}_1$ & $\frac{1}{4}\:-\:\frac{1}{4} \hat{r}^2$ \\
\hline
$\bar{p}_2$ & $\frac{1}{64}\,
\left[12 \omega+15\:-\:(8 \omega+12) \hat{r}^2\:-\:(4 \omega + 3)
\hat{r}^4
\right]$ \\
\hline
$\bar{p}_3$ & $\frac{13}{48}+\frac{19}{48}\omega+\frac{5}{32}\omega^2\:-\:
\frac{1}{256}(53+74 \omega+24 \omega^2) \hat{r}^2 \:-\: 
\frac{1}{256}(9+10 \omega+8 \omega^2) \hat{r}^4 \:-\:
\frac{1}{384}(11+26 \omega + 12 \omega^2) \hat{r}^6$ \\
\hline
$\bar{p}_4$ & $\frac{1}{24576} 
(8593+17719 \omega+12860 \omega^2+3360 \omega^3)\:-\:
\frac{1}{768}(201+413 \omega + 278 \omega^2+60 \omega^3) \hat{r}^2$ \\
 & $-\:\frac{1}{4096}(159+218 \omega+208 \omega^2+96 \omega^3) \hat{r}^4 \:-\:
\frac{1}{1536}(44+93 \omega + 64 \omega^2+24 \omega^3) \hat{r}^6$ \\
 & $-\:
\frac{1}{24576}(503+1707 \omega + 1692 \omega^2+480 \omega^3) \hat{r}^8$ \\
\hline\hline
$C_\xi^{(1)}$ & $-\frac{1}{2}$ \\
\hline
$C_\xi^{(2)}$ & $-\frac{1}{16}(5+6 \omega)$ \\
\hline
$C_\xi^{(3)}$ & $-\frac{1}{128}(37+70\omega+40\omega^2)$ \\
\hline
$C_\xi^{(4)}$ &
$-\frac{1}{3072}(1016+2467\omega+2284\omega^2+840\omega^3)$ \\
\hline\hline
\end{tabular}
\caption{Solution of Eqs.~(\protect\ref{sieq}) up to order $\kappa^4$.
The boundary conditions are described in section IV}
\label{tab:sol}
\end{table*}

\section{\label{sec:ng} Final comments}

We have seen that BDT provides a RTG in D dimensions.
In dimension lower than four, EGR is not a RTG, and the new ingredient 
present in BDT, the scalar field, is essential to have a Newtonian
regime at low energies. The dynamics induced by the new field contains 
\textit{all} features that one expects in a theory of gravitation, so that it
can be used to check phenomena such as gravitational instabilities and
black holes in the simplified context of lower dimensional worlds.

In four dimensions, it is believed that a consistent quantum theory of 
gravitation
must contain, besides the metric tensor, additional fields like the
dilaton. It is remarkable that in lower dimensions the necessity for
such new ingredients appear already at an earlier stage. 

This work has been partially supported
by MCYT (Spain), grants FPA2000-1252 and FPA2001-1813. V.L. has been 
supported by the Ram\'on y Cajal program.

\end{document}